\documentclass[12pt,a4paper]{article}
\usepackage{amsmath}
\input epsf
\usepackage[dvips]{graphicx}
\usepackage{times}

\begin{document}
\begin{titlepage}
\title{Deep--elastic scattering and asymptotics}
\author{ S.M. Troshin,
  N.E. Tyurin\\[1ex] \small\it Institute
\small\it for High Energy Physics,\\\small\it Protvino, Moscow Region, 142281 Russia}
\date{}
\maketitle
\begin{abstract}
Deep--elastic scattering  and its role in  discrimination  of the possible absorptive and reflective asymptotic scattering mechanisms  are discussed with emphasis on
the difference in the experimental signatures related to production processes.

\end{abstract}
\vfill
\end{titlepage}

\section*{Introduction}
The increased interest in  asymptotic scattering mechanism is mainly related to the start of the LHC
wide experimental program. On the theoretical side this interest is also due to the new upper bound for the inelastic cross--section
obtained recently by Andre Martin \cite{am}. Saturation of this bound implies that the limiting form of the partial amplitude of elastic scattering is purely
imaginary one and its modulus is equal to 1/2. Thus, the simultaneous saturation of this new bound and the well-known Froissart--Martin bound for the total
cross-section is excluded since the saturation of the latter one implies that the magnitude of the partial amplitude reaches its maximum value allowed
by unitarity condition, i.e. unity.

The commonly accepted point of view is that the upper limit for the partial amplitude corresponds to the black disk saturation,
i.e. $|f_l|=1/2$. Unitarity limit $|f_l|=1$ is correspondingly discarded\footnote{It should be noted that the above values of the partial amplitudes
 are relevant for the region $l\leq l_{max}(s)$ ($l_{max}(s)=Cs^{1/2}\ln s,\,C\leq \pi/m^2_{\pi}$ \cite{lm}).
Popular ansatz  on the pion mass replacement   to the mass of the heavier particle in the $\pi/m^2_{\pi}\ln^2s$ bound for the total cross-sections
 means change of the value of  $ l_{max}(s)$ and
does not imply any changes of the limitation for  $|f_l|$  in the region $l\leq l_{max}(s)$.}. In the most extreme form this belief was expressed in the recent paper
by Block and Halzen \cite{bh}. In their
paper a rather strong statement was made on the apparent experimental confirmation of the black disk asymptotic mechanism.  It was proposed to use
agreement of the extrapolation based on the particular model parametrization with the new experimental data obtained in the Pierre Auger Observatory \cite{po}
as an argument for the confirmation of the black disk asymptotic regime. However, this result being evidently a model--dependent  cannot
disprove or confirm  the particular models, it is only about their consistency or its absence.  Indeed, using such a simple device as a ruler,
one can immediately arrive
to conclusion that $\ln s$ dependence is in excellent agreement with the experimental data on the inelastic cross-section (cf. Fig. 1 in ref. \cite{bh}).
Then, if $\ln^2 s$--dependence is used for the description of asymptotic dependence of the total cross-section, then one would get zero for the
ratio $\sigma_{inel}(s)/\sigma_{tot}(s)$ at $s\to \infty$. Thus, the existing experimental accelerator and cosmic rays data
cannot constitute  a sufficient set  to make a qualitative conclusion on the possible asymptotic hadron scattering mechanism. This qualitative
conclusion is in line with the recent quantitavive analysis performed  in \cite{mn}.
The true asymptotics in hadron scattering remains therefore to be obscured. The discussion of this problem has a long story and has been performed in
many papers, as an example we would like to note  the two  references \cite{bd} and \cite{js}.

Under such circumstances one can try to  search   the independent indirect qualitative experimental signatures of the possible asymptotic mechanism.
In this connection it is instrumental to consider a deep--elastic scattering.
The notion of deep--elastic  scattering was introduced in the paper \cite{is}.  It is based on analogy with the
deep-inelastic scattering and refers to elastic scattering with the large transferred momenta $-t>4$ (GeV/c)$^2$. This kinematic region is associated with
interactions at small values of  the initial impact parameter of the extended colliding particles (we  consider protons), usually less
than $0.1$ Fermi. Differential cross section of the elastic scattering in this region is commonly described by the power-like
dependence, $\sim (-t)^{-n}$. This comparatively slow decrease\footnote{At low momentum transfers the differential cross--section of elastic scattering
decreases exponentially.} is in agreement with the old and new experimental data and demonstrates significance of elastic channel of reaction
in  this kinematic region.
Such power--like dependence is usually treated  as a manifestation of the QCD structure of a proton (cf.  \cite{mm,bf,ls}).
It should be noted however that this is not a unique interpretation of the power-like behaviour of the elastic scattering cross-sections.
Long time ago such a dependence has been obtained by Serber in the optical model with Yukawa potential. It has also been
obtained in \cite{tt} using analytical properties of the scattering amplitude.

It is evident that all physical processes  and therefore models for their description should conserve probability,
i.e. the corresponding amplitudes should
obey requirements which follow from unitarity of the scattering matrix. In this note we consider deep--elastic  proton scattering and  conclusions
following from unitarity and its saturation.

\section{The two solutions of unitarity for the elastic scattering amplitude}

Unitarity  is formulated
for  the asymptotic colorless  on-mass shell states and therefore it does not   impose direct  constraints
on the  fundamental fields of QCD --- colored fields of quarks and gluons.

 The unitarity condition for the elastic scattering amplitude $F(s,t)$
 can be written in the form
 \begin{equation}\label{un}
 \mbox{Im}F(s,t)=H_{el} (s,t)+H_{inel} (s,t),
 \end{equation}
 where $H_{el,inel}(s,t)$ are the corresponding elastic  and inelastic overlap functions
 introduced by Van Hove \cite{vh}.  This condition interrelates elastic scattering with multiparticle production
processes. In the framework of the probabilistic interpretation of the elastic scattering which was introduced
and discussed in \cite{vp,mk,kn,lk} this relation means presence of the  correlations of the elastic scattering
 with the production processes of various multiplicities.

The functions $H_{el,inel}(s,t)$ are   the Fourier-Bessel transforms of the functions
$h_{el,inel}(s,b)$, i.e.
\begin{equation}\label{hel}
H_{el,inel} (s,t)=\frac{s}{\pi^2}\int_{0}^{\infty} bdb h_{el,inel}(s,b) J_0(b\sqrt{-t}).
\end{equation}
It should be noted that in the impact parameter representation unitarity for the elastic scattering amplitude
has the following form:
\begin{equation}\label{ul}
\mbox{Im} f(s,b)=|f(s,b)|^2+h_{inel}(s,b),
\end{equation}
where $h_{el}(s,b)\equiv |f(s,b)|^2 $ and elastic $S$-matrix element ($2\to 2$) is related to the amplitude as
\begin{equation}\label{sl}
S(s,b)=1+2if(s,b)
\end{equation}

Now we assume the pure imaginary nature of elastic scattering amplitude at high energies. Then, unitarity condition appears to
be a quadratic equation with two roots:
\begin{equation}\label{r1}
 f(s,b)=\frac{i}{2}(1-\sqrt{1-4h_{inel}(s,b)})
\end{equation}
and
\begin{equation}\label{r2}
 f(s,b)=\frac{i}{2}(1+\sqrt{1-4h_{inel}(s,b)}).
\end{equation}
The function $S(s,b)$ becomes real and has a physical interpretation as a survival amplitude of the prompt elastic channel,
i.e. it is the amplitude of probability
that target and projectile particles remain unexcited (survive) during scattering \cite{ch}.
The corresponding expressions for the survival amplitude $S(s,b)$ are the following
\begin{equation}
 S(s,b)=\pm \sqrt{1-4h_{inel}(s,b)},
\end{equation}
i.e. the probability of absorptive (destructive) collisions is $1-S^2(s,b)=4h_{inel}(s,b)$ with $h_{inel}(s,b)\leq 1/4$.
Since we should observe simultaneous vanishing of elastic and inelastic scattering amplitudes at $b\to\infty$,
only one root (\ref{r1}) being usually taken into account, while another one (\ref{r2}) is omitted as a rule. It is
well known shadow approach to elastic scattering. This is valid   in the case when $h_{inel}(s,b)$ is
a monotonically decreasing function of impact parameter and reaches its maximum value at $b=0$.

Now the comment on the  shadow approach should be made. In this approach, the adopted scattering picture evolves with energy towards saturation
of  the black disk limit. In this regime, inelastic overlap function in  impact parameter representation has a central profile and reaches its maximum value,
i.e. $h_{inel}(s,b=0)\to 1/4$ at $s\to\infty$. Thus the survival amplitude described avove  vanishes in the high energy limit
in central hadron collisions, $S(s,b=0)\to 0$.  It should be noted that
the complete
absorption in head-on collisions  does not follow
from unitarity itself and is merely  a result of the assumed
saturation of the black disk limit.

The saturation of the black disk limit means that deep--elastic scattering then  has a completely absorptive nature, i.e. it is strongly correlated
with a maximal probability of the multiparticle production processes. In this relation, it should be noted, that the mean
multiplicity $\langle n\rangle (s)$ can be calculated according the formula
\begin{equation}\label{mm}
\langle n\rangle (s)= \frac{\int_0^\infty  \langle n\rangle  (s,b)h_{inel}(s,b)bdb}{\int_0^\infty h_{inel}(s,b)bdb}.
\end{equation}
In this case the main contribution to $\langle n\rangle (s)$ comes from the region of the small impact parameters.

The deep-elastic scattering should  be relatively  suppressed
 in the regime of the black disk saturation.  It is appropriate to consider the  inelasticity parameter defined
as the ratio of the energy going to the inelastic processes to the total energy. With the simple ansatz of the geometrical
models where inelasticity parameter $K(s,b)$ is proportional to the inelastic overlap function, $K(s,b)=4h_{inel}(s,b)$, the following expression for the
inelasticity averaged over impact distance has been obtained  \cite{ds,bs}:
\begin{equation}\label{in}
 \langle K\rangle(s)=4\frac{\sigma_{el}(s)}{\sigma_{tot}(s)}\left(1-\frac{\sigma_{el}(s)}{\sigma_{tot}(s)}\right).
\end{equation}
From Eq. (\ref{in}) it immediately follows that $1-\langle K\rangle(s)\to 0$ if the saturation of the black disk limit takes
place at $s\to\infty$.
Thus, when the black disk regime is realized, the deep--elastic scattering would be strongly suppressed in favor of
the extensive multiparticle production. Those  events constitute an essential ``competition'' for the deep elastic scattering and
have an isotropic distribution over the asimuthal angle due to the symmetry of the
head-on proton collisions.

\section{Deep--elastic scattering due to reflection}
In this  section we pursue the case of the non-monotonically dependent  inelastic overlap function. It might happen that due to the 
self-damping of the inelastic channels \cite{bbla}
 the  inelastic overlap function $h_{inel}(s,b)$ 
has a peripheral dependence on the impact parameter and vanishes  at $b=0$ in the
high energy limit $s\to\infty$.  
Such behaviour of the function  $h_{inel}(s,b)$ results from the unitarity saturation for the elastic amplitude when
$|f(s,b)|\to 1$ at $s\to\infty$ and $b=0$.  In its turn unitarity saturation straightforwardly follows from the rational form of the unitarization (cf. e.g. \cite{ij}). 
This form
of unitarization might have relation to the confinement property of QCD \cite{cf} and can be used for a qualitative explanation of the regularity
observed in the energy spectrum of the cosmic rays  \cite{cr}. We will return to the rational unitarization form and give some arguments in its favor
later.

Thus, in this case the inelastic overlap function reaches then its maximum value at $b=R(s)$.  Therefore,
 we should take into account the solution of unitarity equation (\ref{r2}),
when $1/2<|f(s,b)|<1$ and $S(s,b)=-\sqrt{1-4h_{inel}(s,b)}$. The scattering starts to be reflective at very
 high energies and  $b=0$. It approaches to the completely
 reflecting limit ($S=-1$) at $s\to\infty$ since $h_{inel}(s,b=0)\to 0$.
The probability of reflective scattering at $b<R(s)$  is determined by the magnitude
 of $|S(s,b)|^2$.

The deep--elastic scattering can be realized as reflective scattering when
colliding hadrons retain their identity in the scattering process and do not suffer from absorption.
The deep--elastic scattering in this approach dominates over multiparticle production since at small impact parameter values
 $h_{inel}(s,b=0)\to 0$.  This provides favorable conditions for the experimental measurements, since it is evident that the peripheral
 profile of $h_{inel}(s,b)$,  associated with reflective scattering,
suppresses the region of small
impact parameters and the main contribution to the mean multiplicity is due to
the peripheral region of $b\sim R(s)$. Deep--elastic scattering in this case is correlated with inelastic events having low cross--section,
i.e. it has a small background due to production and high experimental visibility.

The phenomena that appears here during consideration of scattering at different impact parameter values in the regions $b<R(s)$
and  $b>R(s)$ is similar with appearance of the geometric Berry phase \cite{ij, br}.

The scattering picture described above can be naturally realized
in the $U$--matrix form of unitarization.
In the $U$--matrix approach,
 the survival elastic amplitude $S(s,b)$ is represented in
the following linear fractional transform (we consider pure imaginary amplitude):
\begin{equation}
S(s,b)=[1-U(s,b)]/[1+U(s,b)]. \label{um}
\end{equation}
 $U(s,b)$ is the generalized reaction matrix element (real functions), which is considered to be an
input dynamical quantity. The relation (\ref{um}) is one-to-one transform and easily
invertible. Rational or $U$-matrix form of unitarization was proposed in the theory of radiation dumping  \cite{heit}.
It is closely related to the self-damping of  the inelastic channels  \cite{bbla} and
for the  relativistic case such form of unitarization was obtained in \cite{umat}. 
It  can also be used for construction of a  ``bridge''
between the physical states of hadrons and the states of the confined colour objects \cite{cf}.

Inelastic overlap function $h_{inel}(s,b)$
is connected with $U(s,b)$ by the relation
\begin{equation}\label{hiu}
h_{inel}(s,b)={U(s,b)}/{[1+U(s,b)]^{2}},
\end{equation}
and the only condition to obey unitarity
 is $ U(s,b)\geq 0$. Elastic overlap function is related to the function
 $U(s,b)$ as follows
\begin{equation}\label{heu}
h_{el}(s,b)={U^2(s,b)}/{[1+U(s,b)]^{2}}.
\end{equation}
The form of $U(s,b)$ depends on the particular model assumptions,
but for our qualitative
 purposes it
is sufficient that it increases with energy in a power-like way
 and decreases with impact parameter
like a linear exponent.
At sufficiently  high energies ($s>s_0$),
two separate  regions of
 impact parameters can be anticipated, namely the outer region
of peripheral collisions where the scattering has a typical absorptive origin, i.e.
$S(s,b)|_{b>R(s)}>0$ and
 the inner region of central collisions
where the scattering has a combined reflective and absorptive origin, $S(s,b)|_{b< R(s)}<0$.
 The transition to the negative
values of $S$ leads to
appearance of the real part of the phase shift, i.e. $\delta_R(s,b)|_{b< R(s)}=
\pi/2$ \cite{ij}.
At such high energies (the LHC energies are in this region) the inelastic overlap function
 $h_{inel}(s,b)$ would have a peripheral $b$-dependence and will
tend to zero for $b=0$ at $s\to\infty$ (cf. e.g \cite{ij}).
 In central collisions, $b=0$, an elastic scattering
  approaches to the completely
 reflecting limit $S=-1$ at $s\to\infty$ with vanishing probability of absorptive scattering
$1-S^2(s,b=0)\to 0$. At the same time, the parameter of inelasticity  $\langle K\rangle (s)$ decreases with energy,
$\langle K\rangle (s)\sim 1/\ln s$ at large values of $s$ (cf. (\ref{in})).

The dependence of $R(s)$
 is determined  by the logarithmic function,  $R(s) \sim \frac{1}{M}\ln s $ . It
 is consistent with analytical properties of the resulting elastic  amplitude in
  the complex $t$-plane and mass $M$ can be related to the pion mass.
Thus, the reflective scattering will simulate presence of the repulsive core in
the hadron and meson interactions  at the energies $s> s_R$  and the deep-elastic  scattering will become a dominating process over
corresponding inelastic collisions.
at sufficiently high energies.  Appearance of the reflective
scattering at very high energies might be used for the qualitative explanation of the ``knee'' in the cosmic ray energy spectrum \cite{cr}.

It should be noted that there is a simple approximate relation between cross--sections of reflective $(rl)$ and absorptive $(ab)$
mechanisms of the deep--elastic scattering at the asymptotically high energies, namely
\[
 d\sigma^{(rl)}_{deepel}/dt\simeq  4d\sigma^{(ab)}_{deepel}/dt.
\]
 It follows from the different maximal values of the amplitudes of the head--on hadron collisions, i.e unity --- for the reflective scattering and
1/2 --- for the absorptive scattering mechanisms.

\section*{Discussion and conclusion}

We propose the elastic scattering at large transferred momenta (deep--elastic scattering)  at the LHC as a tool for discrimination
of the asymptotic mechanisms of the hadron scattering. This deep-elastic scattering probes  the region of small impact parameters.
In this region the reflective and absorptive scattering mechanisms have  the most significant differences at high energies.
Those are associated with the different
impact parameter dependencies (profiles) of the inelastic overlap function connected to the particle production.
As a result, the reflective  mechanism  associated with the
complete unitarity saturation will asymptotically decouple from particle production, i.e. at finite energies
 it corresponds to observation of the deep--elastic scattering
 with   decreasing correlations  with   particle production.

Contrary, we emphasize that saturation of the black disk limit
implies strong correlation of deep--elastic scattering with particle production processes.
Respective asymptotic cross--section $d\sigma^{(ab)}_{deepel}/dt$
is expected to be four times lower than in the case of the reflective scattering mechanism.


\small \begin{thebibliography}{99}


\bibitem{am}
A. Martin, Phys. Rev. {\bf{D80}}  (2009)  065013.
\bibitem{lm}
L. Lukaszuk, M. Martin, Nuov. Cim. {\bf{52}} (1967)  122.
\bibitem{bh}
M.M. Block, F. Halzen, Phys. Rev. Lett. {\bf 107}  (2011) 212002.
\bibitem{mn} D.A. Fagundes, M.J. Menon,
arXiv:1112.5115.
\bibitem{bd}
S.M. Troshin, N.E. Tyurin, Phys. Lett. {\bf{B316}} (1993) 175. 
\bibitem{js}
P. Desgrolard, L. Jenkovszky, B. Struminsky, Eur. Phys. J. {\bf{C11}}  (1999)   145.
\bibitem{po}
The Pierre Auger Collaboration, arXiv: 1107.4804.
\bibitem{is}
M.M. Islam, J. Kaspar, R.J. Luddy, Mod. Phys. Lett. {\bf{A24}}  (2009)  485.
\bibitem{mm}
V.A.  Matveev, R.M.  Muradyan,  A.N.  Tavkhelidze,  Lett.  Nuovo Cim. {\bf{7}}  (1973)  719. 
\bibitem{bf}
S.J. Brodsky, G.R. Farrar, Phys. Rev. Lett. {\bf{31}}  (1973 ) 1153. 
\bibitem{ls}
A. Donnachie, P.V. Landshoff, Phys. Lett. {\bf{B387}}  (1996)  637.
\bibitem{tt}
S.M. Troshin, N.E. Tyurin, Fiz. Elem. Chast. Atom. Yadra, {\bf{15}}  (1984)  53.
\bibitem{vh}
L. Van Hove, Rev. Mod. Phys. {\bf{36}}  (1964) 655.
\bibitem{vp}
L. Van Hove, Nuovo Cimento, {\bf{28}}  (1963) 798.
\bibitem{mk}
M. Namiki, Prog. of Theor. Phys. {\bf{33}} (1965)  92.
\bibitem{kn}
Z. Koba, M. Namiki, Nucl. Phys., {\bf{B8}} (1968)  413.
\bibitem{lk}
A.A. Logunov, O.A. Khrustalev,  Sov. J. Part. Nucl., {\bf{1}} (1970)  39.
\bibitem{ch}
N.P. Chang, Nucl. Phys. {\bf{B66}}, 381 (1973).
\bibitem{ds}
J. Dias de Deus, Phys. Rev. {\bf{D32}}  (1985)  2334. 
\bibitem{bs}
S. Barshay, Y. Ciba, Phys. Lett. {\bf{B167}}   (1985)  449.
\bibitem{bbla}
M. Baker, R. Blankenbecler, { Phys. Rev.}  {\bf 128} (1962) 415. 
\bibitem{ij}
S.M. Troshin, N.E. Tyurin, Int. J. Mod. Phys. {\bf{A22}}   (2007)  4437.
\bibitem{cf}
S.M. Troshin, N.E. Tyurin, Int. J. Mod. Phys. {\bf{A26}}   (2011)  4703.
\bibitem{cr}
S.M. Troshin, N.E. Tyurin,
arXiv: hep-ph/0501018.
\bibitem{br}
M.V. Berry, Proc. Roy. Soc. Lond. {\bf{A392}}  (1984) 45.
\bibitem{heit}
W. Heitler, { Proc. Cambr. Phil. Soc.}, {\bf 37}  (1941) 291. 
\bibitem{umat}
A. A. Logunov, V. I. Savrin, N. E. Tyurin and O. A. Khrustalev, Teor. Mat. Fiz.
 { \bf 6}  (1971) 157.
\end{thebibliography}
\end{document}